\documentstyle[psfig,twocolumn,prl,aps]{revtex}
\newcommand{\ltap}{\mbox{$^{_{\textstyle <}}\!\!\!\!\!_{_{\textstyle \sim}}$}}
\newcommand{\aver}[1]{\mbox{$\textstyle \langle$}#1\mbox{$\textstyle \rangle$}}
\begin{document}
\draft
\title{Space and Time pattern of mid-velocity IMF emission\\
in peripheral heavy-ion collisions at Fermi energies
}
\author{S.~Piantelli, L.~Bidini, G.~Poggi, M.~Bini, G.~Casini,
P.R.~Maurenzig, A.~Olmi, G.~Pasquali,\\
 A.A.~Stefanini and N.~Taccetti}
\address{
Istituto Nazionale di Fisica Nucleare and Universit\`a di Firenze,
I-50125 Florence, Italy
}
\date{\today}
\maketitle
\begin{abstract}
The emission pattern in the $V_{perp} - V_{par}$ plane of Intermediate
Mass Fragments with $Z$=3-7 (IMF) has been studied in the collision
$^{116}$Sn + $^{93}$Nb at 29.5 $A$MeV as a function of the Total
Kinetic Energy Loss of the reaction. 
This pattern shows that for peripheral reactions most of IMF's are
emitted at mid-velocity. 
Coulomb trajectory calculations demonstrate that these IMF's are
produced in the early stages of the reaction and shed light on
geometrical details of these emissions, suggesting that the IMF's
originate both from the neck and the surface of the interacting
nuclei. 
\end{abstract}

\narrowtext
An intense emission of intermediate mass fragments (IMF) at
mid-velocity has been put into evidence  
\cite{Bowman93,Montoya94,Toke95,Dempsey96,Toke96,Lukasik97,Plagnol99}
in collisions of heavy ions in the {\it Fermi regime}, 
i.e. at bombarding energies from 30 to $50\;A{\rm MeV}$.
For non-central collisions, where the binary character of the reaction
is preserved, the mid-velocity particles are easily identified in the
$V_{perp} - V_{par}$ plots, the velocity components being usually
defined with respect to the separation axis of the two heavy reaction
partners. 
The velocity region between target and projectile presents an
intensity of particles unexpected on the basis of the statistical
emission from the hot, fully accelerated main fragments.  
The multiplicity of mid-velocity particles is usually obtained by
subtracting from the total emission the contribution of particles
statistically emitted by the projectile- and target-like fragments
(PLF and TLF), see \cite{Lukasik97,Plagnol99}. 
Presently, the isospin composition of the mid-velocity light charged
particles ($Z \leq$ 2, LCP) and IMF's is highly debated, as it can be 
related to the composition of the emitting source, 
the so-called {\it neck} (see, {\it e.g.}, \cite{Poggi01} and
references therein). 
Equally debated is the issue of the mechanism responsible for these
mid-velocity emissions, namely whether they are of dynamical origin or
due to an exotic statistical process, induced by perturbations of the
Coulomb field \cite{Botvina99}. 

It was shown by the INDRA collaboration~\cite{Plagnol99}
that, for a given bombarding energy, the largest ratio 
(up to $\approx$3) of {\it mid-velocity} to {\it statistical} IMF's is
found in non-central collisions. 
However, due the relatively high thresholds for TLF detection
of the existing $4\pi$ detectors, up to now the very peripheral
collisions could not be reliably selected and the mid-velocity
emission of IMF's was never cleanly isolated from the
accompanying statistical emission, thus preventing a detailed study of 
its characteristics. 
 
In this Letter we present experimental results which, thanks to
an efficient selection of peripheral reactions, make it possible to
clearly isolate the mid-velocity emitted IMF's from the statistical
component.
The obtained selection allows -for the first time- to identify
reactions where almost all IMF's concentrate at mid-velocity
and to carefully study their emission pattern. 
Preliminary results were presented in Ref.~\cite{Poggi01}.

The experiment ({\it Florentine Initiative After Superconducting
Cyclotron Opening}, FIASCO) was performed at 
the {\it Laboratori Nazionali del Sud\ } in Catania (Italy).
Targets of $^{93}$Nb ($\approx$200 $\mu g/cm^2$ thick) were bombarded  
with a 29.5$A$MeV pulsed beam of $^{116}$Sn of 
$\leq$0.1 nA intensity and $\leq$1 ns time resolution.
The setup basically consisted of 24
position-sensitive Parallel Plate Avalanche Detectors 
(PPAD)~\cite{CharityMo1:91,Stefanini95} 
covering $\approx\,70\%$ of the forward solid angle.
They measured, with very low thresholds ($<0.1$ $A$MeV),
impact time and position (FWHM resolution of 600 $ps$ and 3.5 $mm$,
respectively) of heavy reaction products ($A$ $\ge$20).
From the velocity vectors, primary (i.e. pre-evaporative)
quantities were deduced, event-by-event, with an improved version
\cite{Casini_nim} of the Kinematic Coincidence Method (KCM).  

With respect to previous experiments \cite{Casini97,Casini99}, the
setup included 160 phoswich scintillators mounted behind 
most of the PPAD's. They detected LCP's and IMF's with $Z \le$ 20
in $\approx 40\%$ of the forward solid angle 
(plus a reduced sampling in the backward hemisphere).
The phoswiches were two-element modules
(fast plastic BC404 + CsI(Tl)) or three-element ones
(BC404 + slow plastic BC444 + CsI(Tl)), 
coupled to fast phototubes.
The fast scintillators had been carefully machined down to $200\;\mu m$ 
in our workshop, with thickness uniformity better than $5\%$ as
required for Z identification up to $Z\approx$ 20;
the thickness of the BC444 was $5\;mm$ and that of the  
CsI(Tl) 30 or 50 $mm$. 
In the three-element modules, the lack of dead layers and the good
optical transmission (obtained by coupling the two plastic elements
with the heat-pressing technique) allowed to identify IMF's with
low thresholds ($\approx$3-10 $A$MeV for $Z$=2-20).
The measurement of the time of flight allowed to directly obtain
the velocities of LCP's and IMF's, without time
consuming and tricky energy calibrations of the scintillators.
A detailed description of the apparatus can be found in
\cite{tesisilvia}.

The data presented in this Letter are focused on binary events, where
only two major fragments are detected in the PPAD's while the
coincident LCP's and IMF's hit the phoswich detectors. 
Using the Total Kinetic Energy Loss (TKEL) as an ordering variable, it
is possible to select events with increasing impact parameter up to
grazing collisions.
At variance with respect to all available $4\pi$ detectors, our
low-threshold apparatus makes it possible to measure both the PLF and
the TLF also for events with TKEL$\leq$ 400 MeV.
Reconstructed primary quantities were obtained from the KCM with
2-body kinematics, not only for LCP, but also for IMF emission
\cite{notaKCM}.

\begin{figure}[t]
\centerline{\psfig{figure=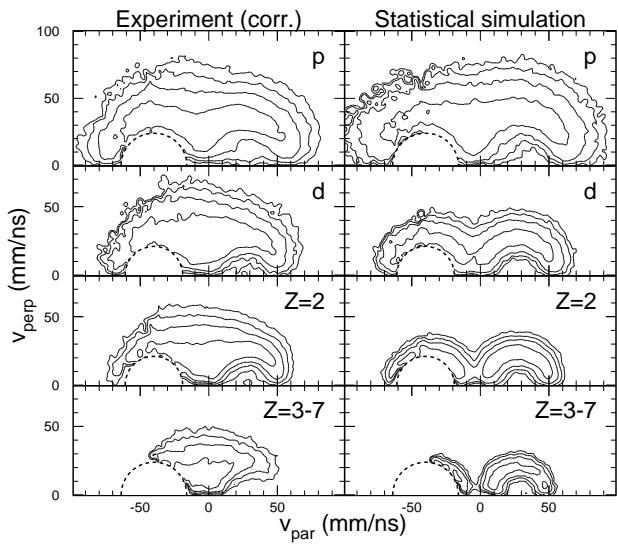,width=82mm,clip=t,%
    bbllx=5bp,bblly=35bp,bburx=520bp,bbury=500bp}}
\caption{
Left: experimental yields in the $V_{perp}$-$V_{par}$ 
plane for p, d, He and IMF's ($Z=3-7$) in the system
$^{116}$Sn +  $^{93}$Nb at 29.5 $A$MeV, for TKEL =240-400 MeV
(corrected for the setup geometry).
Level spacing is logarithmic, dashed lines indicate velocity
thresholds. 
Right: same results for the simulated statistical
emission from the hot reaction partners.
}
\label{occhi}
\end{figure}

The left panels of Fig.~\ref{occhi} present, for TKEL=240-400 MeV,
the experimental yield of p, d, He and IMF
($Z$=3-7) in the $V_{perp}$ - $V_{par}$ plane, 
$V_{par}$ and $V_{perp}$ being the center-of-mass velocity components
parallel and perpendicular, respectively, to the separation axis of
the two major fragments. 
The dotted lines show the velocity thresholds due to the 
thickness of the thin plastic scintillator. 
The data have been corrected for the finite geometry of the
apparatus \cite{tesisilvia};
because of its large acceptance and axial symmetry,
the correction is largely independent of the emission pattern.
In the backward lab-hemisphere ($V_{par} \ltap -40\;mm/ns$), owing to
the reduced detector coverage, the correction is not as effective 
as forwards.

In order to better put into evidence the major features of the
experimental data, the right panels of Fig.~\ref{occhi} show
the corresponding yields obtained with the simulation of a pure
statistical emission from fully accelerated fragments,
filtered with the setup acceptance and then corrected as the
experimental data.
The excitation energy of the fragments was obtained from TKEL
with an ``equal energy'' sharing, according to \cite{Casini00}, 
while first guesses of the parameters of the evaporation step were
deduced from calculations with the code Gemini \cite{Charity88}.

The experimental emission pattern for protons does not differ
very much from that expected for a sequential evaporation, but with
increasing particle mass the mid-velocity yield becomes increasingly
important until it actually exhausts most of the IMF intensity.  

The mid-velocity particle multiplicities were obtained with a
procedure similar to that outlined in \cite{Lukasik97}.
The simulation was tuned so as to reproduce the experimental data in
the velocity region corresponding to forward emission from the PLF
($10^{\circ} \le \theta_{PLF} \le 40^{\circ}$ in the PLF reference frame). 
Assuming that the simulation properly mimics the whole statistical
emission for PLF and TLF, the mid-velocity yield is obtained as the
difference between the total experimental emission and the
corresponding estimation of the statistical component. 

\begin{figure}[t]
\centerline{\psfig{figure=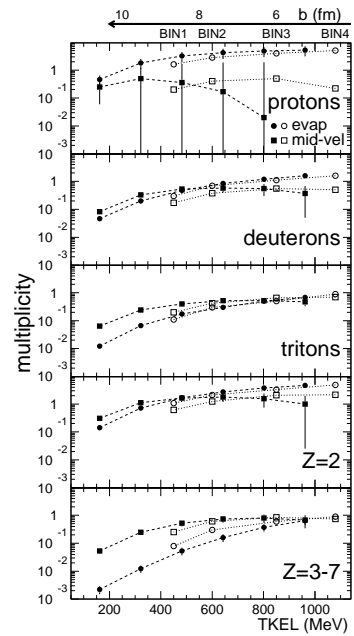,width=50mm,clip=t,%
   bbllx=0bp,bblly=0bp,bburx=400bp,bbury=680bp}}
\caption{
Experimental multiplicities of p, d, t, He and IMF's ($Z$=3-7) 
against TKEL in $^{116}$Sn + $^{93}$Nb at 29.5 $A$MeV.
Full squares (circles) refer to the mid-velocity (evaporative)
component; 
open symbols are for the system Xe + Sn at 32 $A$MeV
\protect\cite{Plagnol99}. 
Lines are to guide the eye.
On top, correspondence between TKEL and impact parameter (or
centrality binning of~\protect\cite{Plagnol99}), estimated from
the QMD code CHIMERA~\protect\cite{Lukasik93}.
}
\label{multiplicity}
\end{figure}

The deduced multiplicities are shown as a function of TKEL in
Fig.~\ref{multiplicity}, separately for p, d, t, He and IMF.
Full circles refer to the statistical evaporation, with uncertainties
(due to detector thresholds and determination of the TLF emission)
usually smaller than the symbol size. 
Full squares show the multiplicity of mid-velocity particles.
The larger the mid-velocity component, the more reliable 
its extraction is;
the largest errors are for protons, owing to the presence of a large
statistical component.

It is worth noting that, in the mid-velocity emission of peripheral
events, the mass removed by the IMF's (assuming $A \approx 2Z$) is
comparable to that removed by LCP's 
(while it is much smaller in case of evaporation).
Moreover, as already qualitatively seen in Fig.~\ref{occhi}, the
mid-velocity component of IMF's greatly overcomes the statistical
emission, by more than a factor of 20 for the most peripheral
collisions. 
To our knowledge, this extremely enhanced emission of mid-velocity
IMF's has never been directly observed before. 
In these events the nuclear matter tends to break apart in
intermediate mass fragments, this process successfully competing with
LCP emission. 

The comparison of the multiplicities presented in
Fig.~\ref{multiplicity} with those of Xe+Sn (open symbols) at similar 
bombarding energies \cite{Plagnol99}, requires to find a
correspondence between our ordering variable (TKEL) and that
(transverse energy of LCP's, E$_{trans12}$) used by the INDRA
collaboration to estimate the impact parameter. 
Following ref.~\cite{Plagnol99}, we used the same QMD code
CHIMERA~\cite{Lukasik93} as an event generator for our reaction
$^{116}$Sn + $^{93}$Nb. 
In simulated events, analyzed as the experimental ones, the
reconstructed TKEL is narrowly correlated with the impact parameter
\cite{note}; details are given in~\cite{tesisilvia}.
Here, on top of Fig.~\ref{multiplicity}, we synthetically draw the
obtained impact parameter scale, the notation
underneath ({\it BIN1, ..., BIN4}) corresponding to the binning
of \cite{Plagnol99}.   
It has to be noted that the range of impact parameters probed by our
experiment extends to significantly larger values than those
accessible with INDRA, the results of both experiments being in good
agreement in the common region.

Addressing the question whether the production mechanism of
mid-velocity IMF's is mainly of dynamical or statistical nature is
beyond the scope of this letter.  
However, the peculiar emission pattern observed in the very peripheral
collisions (where it is almost free from significant contaminations
due to evaporation from PLF and TLF) gives information on the
time-space configuration of the emitting system.
Namely, it is possible to infer the time scale and the geometrical
configuration at the end of the reaction from Coulomb trajectory
calculations reproducing the observed emission pattern. 
Although qualitative descriptions of the IMF emission based on Coulomb
trajectory calculations are found in the literature (see, {\it e.g.}
\cite{Toke95_ta}), quantitative conclusions are not frequent.  

\begin{figure}[t]
\centerline{\psfig{figure=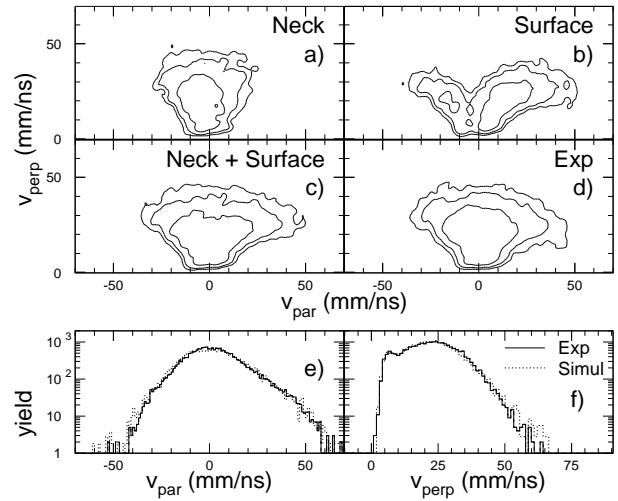,width=82mm,clip=t,%
    bbllx=0bp,bblly=10bp,bburx=520bp,bbury=450bp}}
\caption{
$V_{perp}$-$V_{par}$ plots for IMF's (Z=3-7) at TKEL= 240-400 MeV:
(a) Coulomb trajectory calculation for {\it neck-region emission} with
    $\aver{E_n}$ =25 MeV and 
(b) for {\it surface emission} with $\aver{E_s}$ =10 MeV, 
(c) superposition of the two contributions and 
(d) experimental correlation.
Projection of experimental (full histogram) and simulated (dotted
histogram) yields on (e) the $V_{par}$ and (f) the $V_{perp}$ axis.
}
\label{coulomb}
\end{figure}

A {\it neck-region emission} was assumed, with the IMF's emitted from
a ``neck'' of excited nuclear matter connecting the two interacting
nuclei. 
Schematically, the Coulomb trajectory calculation was implemented in
our Monte Carlo simulation as follows. 
The two main fragments were assumed to be in an initial configuration
(corresponding to the selected peripheral impact parameter) at a
distance given by a tuning parameter ($d_{\rm sep}$) of the calculation.
The to-be-emitted IMF, sampled from a realistic distribution, 
was located in between the two major
fragments, with a random ``thermal-like'' kinetic energy described by
another tuning parameter $\aver{E_n}$.
The initial separation velocity of the two main fragments was that
appropriate for producing the correct asymptotic TKE.
The equations of motion were integrated starting from this simple 
initial configuration in the three-body phase space, then the
experimental filter was applied and the $V_{perp} - V_{par}$ plots
were built.  
Fig.~\ref{coulomb}a presents the obtained emission pattern for impact
parameters corresponding to the experimental data in the bin TKEL=
240-400 MeV. 
The main effect of the two close-lying fragments is to focus the
emitted IMF in the transverse direction, with the shape of the
$V_{perp}$ distribution mainly depending on the assumed
thermal-like energy. 
In fact, for peripheral collisions, the heavy fragments fly swiftly
apart, with no significant energy transfer to the IMF's
if they are emitted nearly at rest in the CM system.
From the comparison with the experimental correlation
(Fig.~\ref{coulomb}d), one sees that the simulated mechanism of 
{\it neck-region emission} significantly 
contributes to the experimentally observed mid-velocity emission, in
particular in the high transverse momentum region, but there is no way
to reproduce the two Coulomb-like wings which in the experimental data 
start at mid-rapidity and rapidly fade away when moving forward
(backward) in the PLF (TLF) emission frame.  

It was therefore assumed that a second contribution exists, similar
to the so-called 
fast oriented fission \cite{Stefanini95,Casini93}
where the very asymmetric fission-like decays of $A\approx 100$
nuclei proceed through nearly aligned configurations
(light fission fragment in between the heavier one and the other
non-fissioning reaction partner) with characteristic times of the
order of $10^{-21}\;s \approx 300\;fm/c$.
We performed Coulomb trajectory calculations assuming a fast
{\it surface emission} of IMF's from the contact regions on the
surface of the two flying-apart primary fragments. 
The IMF was emitted, again with a ``thermal-like'' energy 
$\aver{E_s}$, from a point of the surface randomly spread within a
few {\it fm} from the separation axis.
The emission pattern obtained for such emissions is presented 
in Fig.~\ref{coulomb}b.
This {\it surface} mechanism reproduces the wings of the experimental 
distribution (Fig.~\ref{coulomb}d), but not the large transverse
energies:  
in this case the distribution is mainly determined by the Coulomb
field of the emitting fragment.

In order to satisfactorily reproduce the experimental data, both
contributions must be included. 
The best agreement with the experiment has been obtained for a
ratio between {\it surface} and {\it neck-region emission} of 0.7 and
with values of $\aver{E_s}$= 10 MeV and $\aver{E_s}$= 10 MeV:
the result is shown in Fig.~\ref{coulomb}c.
The projections of the experimental data and the simulation
(continuous and dotted histograms, respectively) on the $V_{par}$
(Fig.~\ref{coulomb}e) and $V_{perp}$ (Fig.~\ref{coulomb}f) axes 
allow to better perceive the good quality of the obtained agreement.
In both projections, the simulation presents a sizeable sensitivity to
the assumed random energies $\aver{E_n}$ and $\aver{E_s}$. 
In fact, a change of 5 MeV in these parameters worsens in a
significant way the agreement.
An agreement of the same quality is found for more peripheral events 
(TKEL$\leq$240 MeV), while for more central TKEL-bins the comparison
becomes increasingly blurred by the growing contribution of the
statistical emission. 
The overall trend is that both the random energy $\aver{E_n}$ of the
{\it neck-region emission} and the spread of the emission point
in the {\it surface emission} increase with increasing centrality. 

For peripheral events, the experimental IMF emission appears to be
compatible with the formation of a neck-like structure which then
decays by a prompt emission from the neck region itself or 
by a successive emission from the surfaces of the separating nuclei.
For example, the {\it neck-region emission} may be due to a multiple 
neck rupture, while the {\it surface emission} is suggestive of a 
single neck rupture \cite{neckmodels}, 
leaving one of the main fragments in a rather
deformed shape, in the vicinity (or even beyond) its saddle shape.
In the simulation these two mechanisms differ in terms of the
initial configuration: the {\it neck-region emission} contributes at
small distances $d_{\rm sep}$ compatible with the presence of a
connecting neck, the {\it surface emission} at larger distances. 
According to models quantitatively describing the formation of the
neck in nuclear collisions or fission \cite{neckmodels}, the maximum
length of the neck --discriminating between the two mechanisms-- 
was estimated to be $12\;fm$.
From the relative kinetic energy at separation and from the envisaged
geometrical configuration, it is possible to estimate the time scale:
a value of the order of 60 $fm/c$ is obtained as a discrimination
between {\it neck-region} and {\it surface emission}.
Moreover, from the observed anisotropy and using realistic values of
the angular momenta and moments of inertia of the nuclei, one can
exclude for the {\it surface emission} times significantly larger than
a few hundreds of $fm/c$.

In conclusion, peripheral collisions are characterized by an important
emission of IMF's at mid-velocity, successfully competing with LCP's.
Indications for a fast energy dissipation process of locally excited
nuclear matter may be inferred from the estimated {\it thermal-like}
energies necessary to reproduce the data.
The emission pattern of IMF's is compatible with the coexistence
of two mechanisms of prompt or fast emission: the first one related to
the formation and prompt break-up of a neck of highly excited nuclear
matter and the second one, at a somewhat later stage, characterized by
a localized emission from the possibly highly deformed flying apart
fragments. 
 
We wish to thank J.~{\L}ukasik for providing
us with the QMD code CHIMERA. 
We are grateful to R. Ciaranfi and M. Montecchi for the development of
dedicated electronics, and to P. Del Carmine for his help in the
setup preparation. 
Many thanks are due to L. Calabretta and to the whole machine crew of the LNS
for their continuous efforts to provide a good quality pulsed beam.


\begin{references}
\bibitem{notaKCM}
It was checked that in case of emission of just one IMF (the average
IMF multiplicity is well below 1 for semiperipheral collisions)
good agreement is found between the results of the analysis with the
2-body kinematics (ignoring the IMF) and with the 3-body kinematics 
(including the IMF as a third body).
\bibitem{note} The same analysis showed that, because of the reduced 
  acceptance of our setup, the $E_{trans12}$ selection gave no narrow
  correlation with the impact parameter.
\end{references}
\end{document}